\documentclass[published]{epl}
\usepackage{graphicx}

\title{Ground state of two unlike charged colloids: An analogy with ionic bonding}
\shorttitle{Ground state of two unlike charged}
\author{R. Messina\thanks{E-mail: \email{messina@mpip-mainz-mpg.de}}
       \and C.Holm
       \and K. Kremer}

\institute{ Max-Planck-Institut f\"ur Polymerforschung, Ackermannweg 10,
          55128 Mainz, Germany}
\issue{51}{4}{2000}{461}{15 August 2000}
\rec{4 April 2000}{22 June 2000}
\begin{document}
\maketitle
%

%%%   ! Don't forget this command to format the title page of your article!

%

%%%   The Abstract

%

\begin{abstract}
In this letter, we study the ground state of two spherical macroions of identical
radius, but asymmetric bare charge (\( Q_{A}>Q_{B} \)). Electroneutrality of
the system is insured by the presence of the surrounding divalent counterions.
Using Molecular Dynamics simulations within the framework of the primitive model,
we show that the ground state of such a system consists of an overcharged and
an undercharged colloid. For a given macroion separation the stability of these
ionized-like states is a function of the difference (\( \sqrt{N_{A}}-\sqrt{N_{B}} \))
of neutralizing counterions \( N_{A} \) and \( N_{B} \). Furthermore the degree
of ionization, or equivalently, the degree of overcharging, is also governed
by the distance separation of the macroions. The natural analogy with ionic
bonding is briefly discussed.
\end{abstract}
%

%

%%%   Main text

%

%%%   Sectioning

%

%%%   In EuroPhys there is only ``one'' level of sectioning `\section{}'.

%

%Main text begins here.

Charged colloids are found in a great variety of materials such as latex, clays,
paints, and many biological systems, and thus have an important place in the
every-day life. To understand the complex interaction between charged colloids
and their surrounding neutralizing counterions, a reasonable starting point
is to study the elementary case of a pair of spherical macroions. From the theoretical
side such a system is described by the Derjaguin-Landau-Verwey-Overbeek (DLVO)
theory \cite{Derjaguin,Verwey}. which lead to purely
repulsive effective forces. More sophisticated modified Poisson-Boltzmann approaches
based on density-functional theory \cite{Groot_JCP_1995} or inhomogeneous
HNC techniques \cite{Belloni_1985,Marcelo_CPL_1992}
have been developed in order to incorporate the ion-ion correlations which are
neglected in DLVO. Surprisingly recent experiments showed effective attractive
forces between like-charged colloids 
\cite{Kepler_1994,Crocker_1996,Larsen_1997}
when they are confined near charged walls, and for which no clear theoretical
explanation is available. This triggered reinvestigations of the pair-interactions
in the bulk with computer simulations 
\cite{Jensen_PhysicaA_1998,Allahyarov_PRL_1998,Allahyarov_PRE_1999,Wu_1999,Messina_PRL_2000}.
A common feature of all these studies is that they assume the two macroions
identically charged. The results of refs. 
\cite{Jensen_PhysicaA_1998,Allahyarov_PRL_1998,Allahyarov_PRE_1999,Wu_1999}
show for high Coulomb coupling an attractive force in a range of the order of
a few counterion radii. However, Messina et al. \cite{Messina_PRL_2000}
have demonstrated that it is possible to get a strong long-range attraction
between two like-charged colloids due to metastable ionized states. In particular
it has been shown that the energy difference between the compensated bare charge
case, where each colloid is exactly neutralized by the surrounding counterions,
and the ionized state can be very small (less than 2 $k_{B}T$).

In this letter, we use molecular dynamics (MD) simulations to investigate the
case where the colloidal radii are identical but the bare colloidal charges
are different. It is found that in this asymmetric situation the ground state
is no longer the intuitive bare charge compensated case, provided that the charge
asymmetry is high enough and/or the colloid separation is not too large. We
derive a simple formula valid for large separations which gives a sufficient
condition for the bare charge asymmetry, to produce a ground state consisting
of an ionic pair leading to a natural long-range attractive force.

The system under consideration is made up of two spheres: (i) macroions $A $
and $ B $) of diameter \textit{d} with bare charges $ Q_{A}=-Z_{A}e $
(where $e$ is the elementary charge and $Z_{A}= 180$ is
\textit{fixed}) for the highly charged sphere and $ Q_{B}=-Z_{B}e $ (\textit{variable})
for the less charged one and (ii) a sufficiently number of small counterions
of diameter $ \sigma  $ with charge $ q=+Z_{c}e $ ($ Z_{c}=2 $)
to neutralize the whole system. The macroions center-center separation is given
by \textit{R}. The ions are confined in a cubic box of length \textit{L}, and
the two macroions are held fixed and disposed symmetrically along the axis passing
by the two centers of opposite faces. The colloid volume fraction \textit{$ f_{m} $}
is defined as $ 2\cdot 4\pi (d/2)^{3}/3L^{3} $. For describing the charge
asymmetry we define the quantity $ \alpha =\sqrt{N_{A}}-\sqrt{N_{B}} $, where
$ N_{A}=-Q_{A}/q $, and $ N_{B}=-Q_{B}/q $.

The motion of the counterions is coupled to a heat bath acting through a weak
stochastic force \textbf{W}(t). The equation of motion of counterion \textit{i}
reads 
%
%%%%%%%%%%%%%%%%%%%%%%%%
\begin{equation}
\label{eq. Langevin}
m\frac{d^{2}\mathbf{r}_{i}}{dt^{2}}=
-\nabla _{i}U-m\Gamma \frac{d\mathbf{r}_{i}}{dt}+\mathbf{W}_{i}(t)\: ,
\end{equation}
%%%%%%%%%%%%%%%%%%%%%%%%
%
where \textit{m} is the counterion mass, $ \Gamma  $ is the friction coefficient,
chosen here between 0.1 and 1.0, and \textit{U} is the potential consisting
of the Coulomb interaction and the excluded volume interaction. Friction and
stochastic force are linked by the fluctuation-dissipation theorem $ \langle {\mathbf{W}_{i}}(t)\cdot {\mathbf{W}_{j}}(t')\rangle =6m\Gamma k_{B}T\delta _{ij}\delta (t-t^{'}) $.
In the ground state $ T=0 $ and thus the stochastic force vanishes.

Excluded volume interactions are taken into account with a pure repulsive Lennard-Jones
(LJ) potential given by 
%
%%%%%%%%%%%%%%%%%%%%%%%%
\begin{equation}
\label{eq. LJ}
U_{LJ}(r)=\left\{ \begin{array}{l}
4\varepsilon \left[ \left( \frac{\sigma }{r-r_{0}}\right) ^{12}-\left( \frac{\sigma }{r-r_{0}}\right) ^{6}\right] +\varepsilon ,\\
0,
\end{array}\qquad \right. \begin{array}{l}
\textrm{for}\, \, r-r_{0}<2^{1/6}\sigma ,\\
\textrm{for}\, \, r-r_{0}\geq 2^{1/6}\sigma ,
\end{array}
\end{equation}
%
%%%%%%%%%%%%%%%%%%%%%%%%
% 
where $ r_{0}=0 $ for the counterion-counterion interaction, $ r_{0}=7\sigma  $
for the macroion-counterion interaction, thus leading to a macroion diameter
$ d=2r_{0}+\sigma  $ and electrostatically more important to a macroion-counterion
distance of closest approach $ a=8\sigma  $.

The pair electrostatic interaction between any pair \textit{ij}, where \textit{i}
and \textit{j} denote either a macroion or a counterion, reads 
%
%%%%%%%%%%%%%%%%%%%%%%%%
\begin{equation}
\label{eq. coulomb}
U_{coul}(r)=k_{B}T_{0}l_{B}\frac{Z_{i}Z_{j}}{r}\: ,
\end{equation}
%
%%%%%%%%%%%%%%%%%%%%%%%%
%
where $ l_{B}=e^{2}/4\pi \epsilon _{0}\epsilon _{r}k_{B}T_{0} $ is the Bjerrum
length describing the electrostatic strength. To link this to experimental units
and room temperature we denote $ \varepsilon = $$ k_{B}T_{0} $ ($ T_{0}=298 $
K). Fixing $ \sigma =3.57 $ \AA\ would then lead to the Bjerrum length of
water at room temperature (7.14 \AA).

Being interested in the strong Coulomb coupling regime we choose the relative
permittivity $ \epsilon _{r}=16 $, corresponding to $ l_{B}=10\sigma  $.

%%%%%%%%%%%%%%%%%%%
% FIG 1
\begin{figure}[b]
\oneimage[width = 6.0 cm]{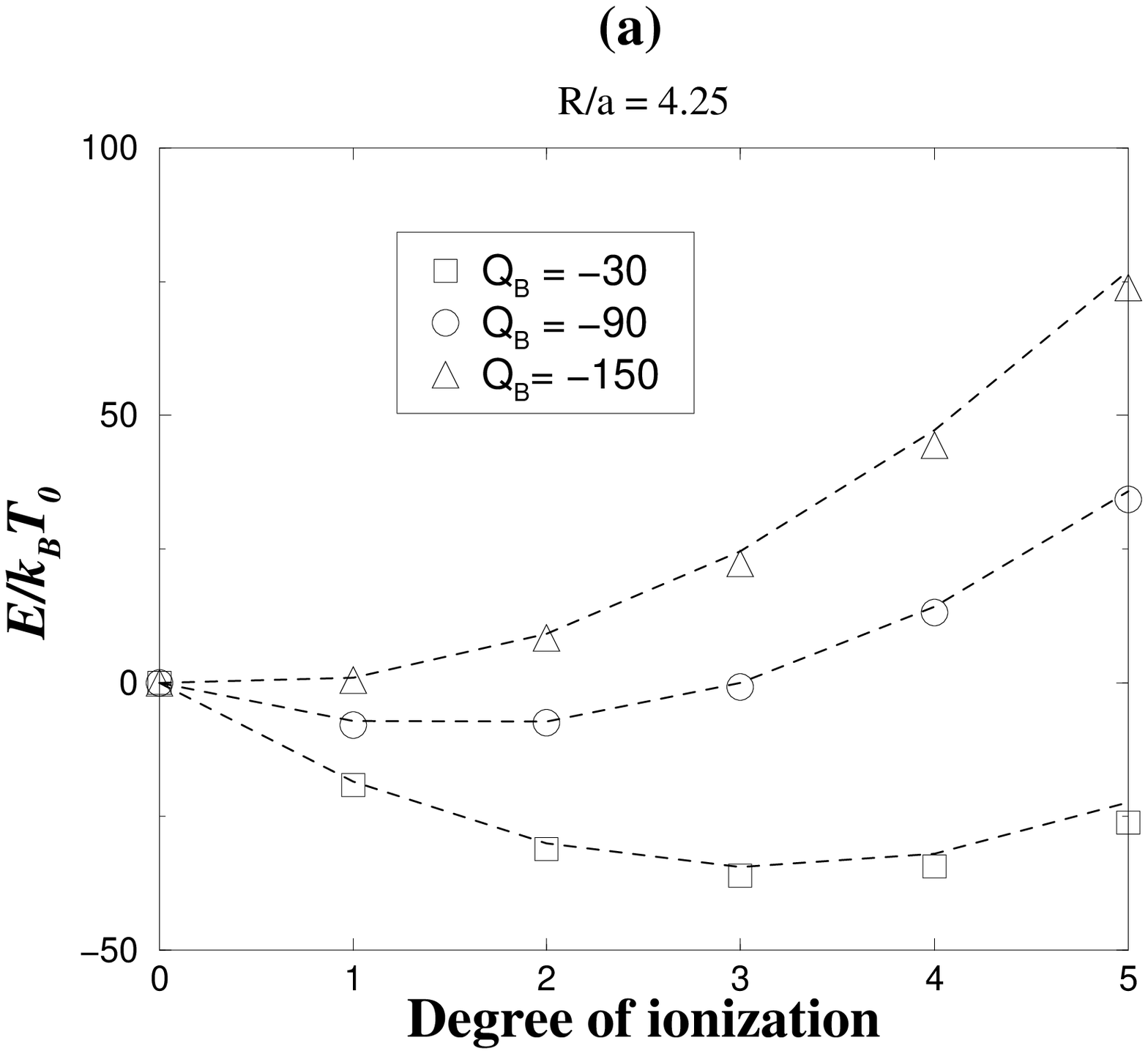}
\twoimages[width = 6.0 cm]{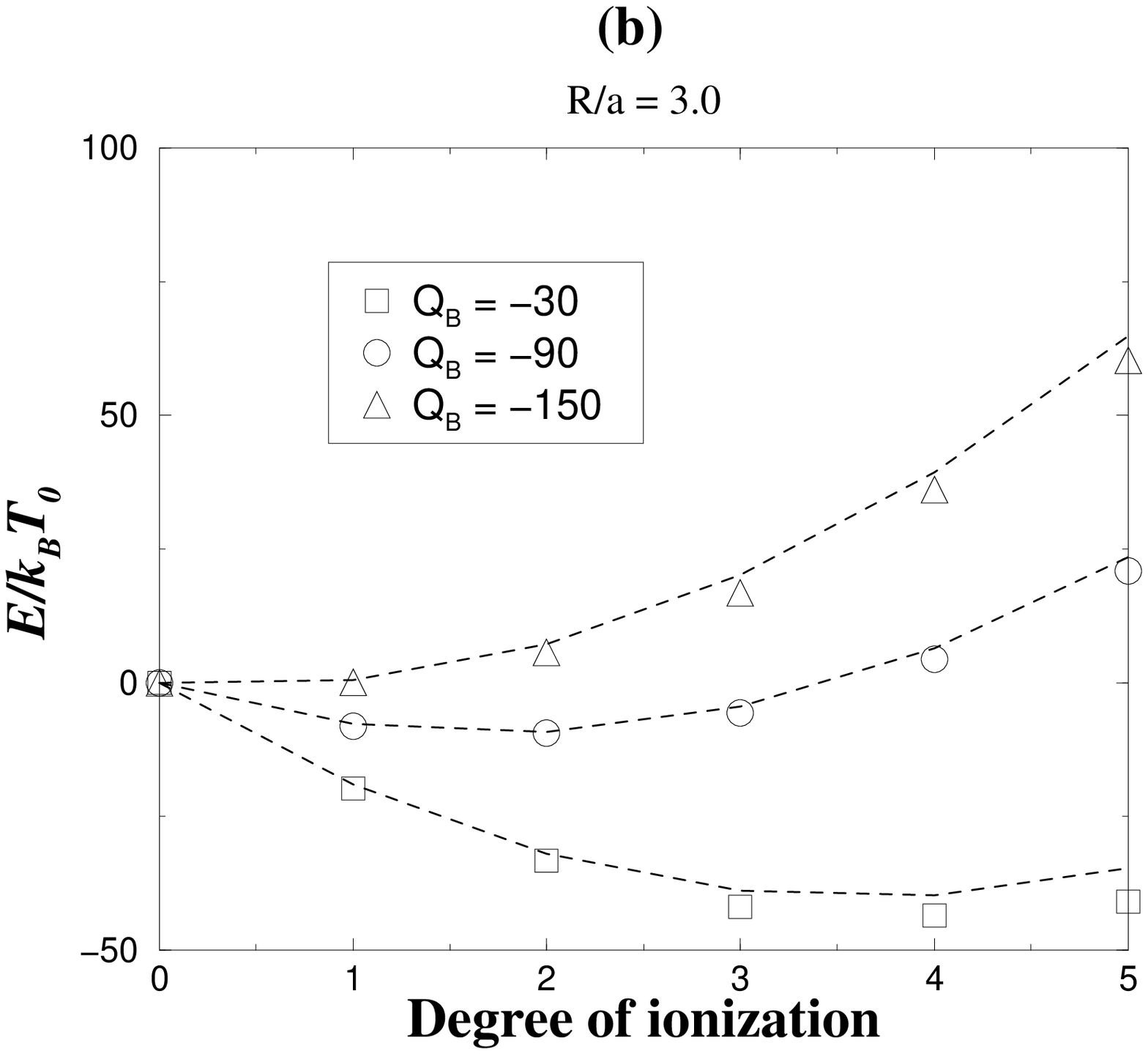}{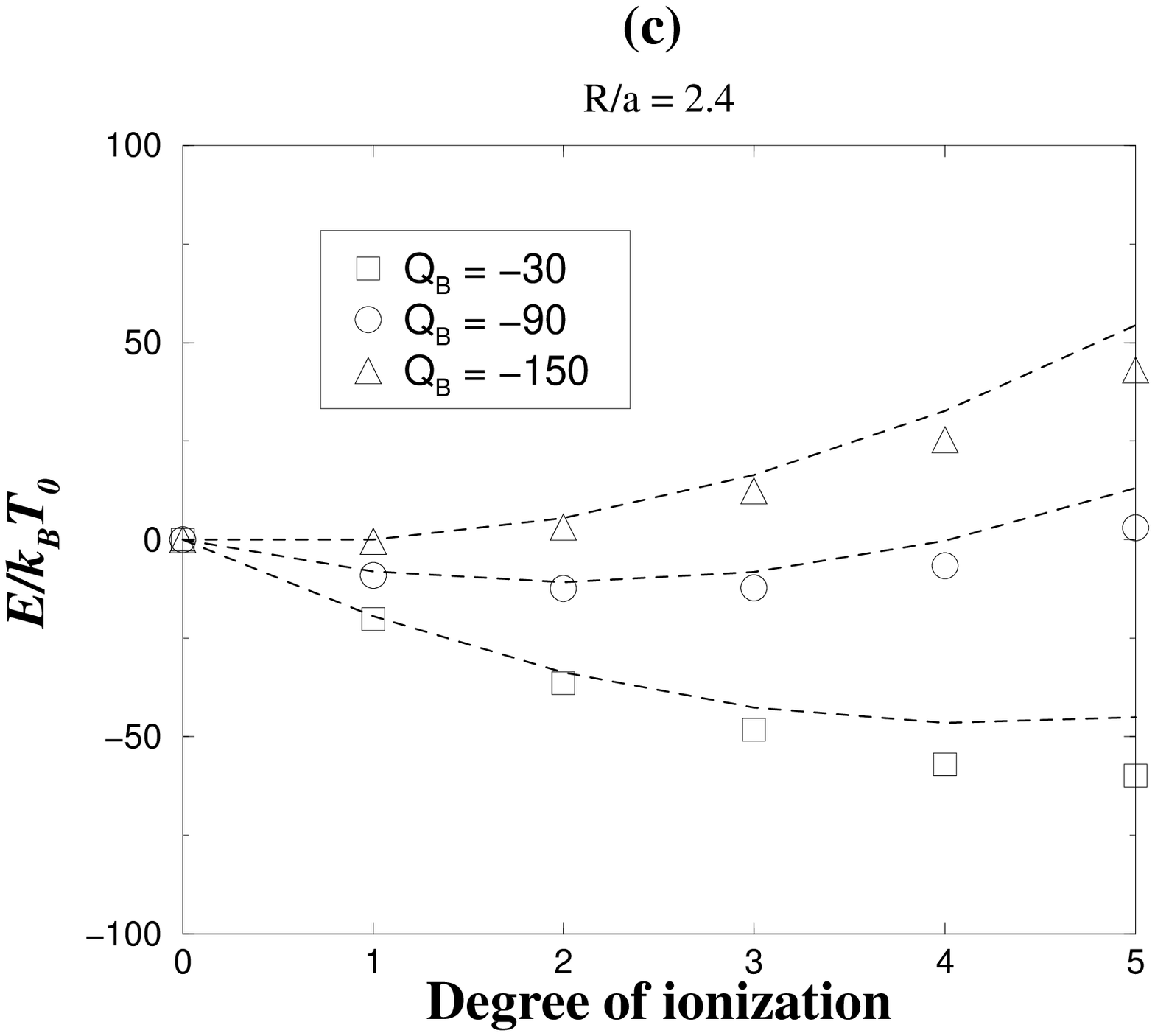}
\caption{
Total electrostatic energy as a function of the degree of ionization
for zero temperature configurations of two colloids ($A$ and $B$),
for three typical charges $Q_B/e$ ($ -30,-90 $ and $-150$) for
macroion $B$ and for three given distance separations: a) $ R/a=4.25 $,
b) $ R/a=3.0 $ and c) $ R/a=2.4 $. Dashed lines are obtained using eq.
\ref{eq.finiteR}.
}
\label{davgol}
\end{figure}
%%%%%%%%%%%%%%%%%%%%%

The electrostatic energy of the system is investigated for different uncompensated
bare charge cases \cite{NOTE_MD} by simply
summing up eq. (\ref{eq. coulomb}) over all Coulomb pairs. Note that for the
zero temperature ground state study entropic effects are nonexistent. We define
the \textit{degree of ionization} (\textit{DI}) as the number of counterions
overcharging colloid A (or, equivalently, undercharging colloid B). The system
is prepared in various \textit{DI} and measure the respective energies. These
states are separated by kinetic energy barriers, as was demonstrated in Ref.
\cite{Messina_PRL_2000}. We consider three typical macroionic charges $ Z_{B} $
(30, 90 and 150) and separations $ R/a $ (2.4, 3.0 and 4.25). The main results
are given in Fig. 1. For the largest separation $ R/a=4.25 $ and largest
charge $ Z_{B}=150 $ (see Fig. 1a), one notices that the ground state corresponds
to the classical compensated bare charge situation (referred as the \textit{neutral}
\textit{state}). Moreover the energy increases stronger than linear with the
degree of ionization. If one diminishes the bare charge $ Z_{B} $ to 90 and
30, the \textit{ground state} is actually the \textit{ionized state} for a \textit{DI}
of 1 and 3, respectively. The ionized ground state is about 8 and 36 $ k_{B}T_{0} $,
respectively, lower in energy compared to the neutral state. This shows that
even for a relative large colloid separation, stable ionized states should exist
for sufficient low temperatures and that their stability is conditioned by the
structural charge asymmetry $ \alpha  $.

For a shorter separation $ R/a=3.0 $, ionized ground states are found (see
Fig.1 b) for the same charges $ Z_{B} $ as previously. Nevertheless, in the
ground state the \textit{DI} is now increased and it corresponds to 2 and 4
for $ Z_{B}=90 $ and 30 respectively. The gain in energy is also significantly
enhanced. For the shortest separation under consideration $ R/a=2.4 $, the
ground state corresponds for \textit{all} investigated values of $ Z_{B} $
to the ionized state, even for $ Z_{B}=150 $. We conclude that decreasing
the macroion separation \textit{R} enhances the DI and the stability of the
ionized state.

To understand this ionization phenomenon, it is sufficient to consider an \textit{isolated}
macroion surrounded by its neutralizing counterions. We have investigated
the energies involved in the ionization (taking out counterions) and overcharging
(adding counterions) processes. We show in Ref. \cite{Messina_PRL_2000} how
they can be separated into two parts: (i) a pure correlational term ($ \Delta E^{cor} $)
and (ii) a monopole contribution ($ \Delta E^{mon} $), see also Ref. \cite{Shklowskii_PRE_1999}
for the case of added salt. The main assumption is that the correlational energy
per ion can be written as a pure surface term $ \epsilon (N)=-\gamma \sqrt{N} $
(with $ \gamma >0) $, as is predicted for example in a theory where the counterions
on the surface of the colloids form a Wigner crystal (WC)
\cite{Shklowskii_PRE_1999,WCT}. 
The gain in energy when adding the first counterion is simply
a pure correlation term of the form 
%
%%%%%%%%%%%%%%%%%%%%%%%%%%%
\begin{equation}
\label{eq.firstoc}
\Delta E^{OC}_{1}=
\Delta E^{cor}_{1}=(N_{A}+1)\epsilon (N_{A}+1)-(N_{A})\epsilon (N_{A})=
-\gamma \sqrt{N_{A}}\left[ \frac{3}{2}+\frac{3}{8N_{A}}+{{\mathcal{O}}}(N_{A}^{-2})\right] \: .
\end{equation}
%%%%%%%%%%%%%%%%%%%%%%%%%%%
%
 Adding the summed up monopole contributions, one obtains the energy gained
by adding the $ n^{th} $ counterion to leading order in $ 1/N_{A} $: 
%
%%%%%%%%%%%%%%%%%%%%%%%%%%%
\begin{equation}
\label{Eq.oc}
\Delta E^{OC}_{n}=\Delta E^{cor}+\Delta E^{mon}=-n\gamma \sqrt{N_{A}}\left[ \frac{3}{2}+\frac{3n}{8N_{A}}\right] +(k_{B}T_{0})l_{B}Z_{c}^{2}\frac{(n-1)n}{2a}\: ,
\end{equation}
%%%%%%%%%%%%%%%%%%%%%%%%%%%
%
which has been verified to give a correct description when compared to simulations
\cite{Messina_PRL_2000}. A derivation of the formula describing the ionization
energy $ \Delta E^{ion} $ proceeds completely analogously and gives for the
$ n^{th} $ degree of ionization 
%
%%%%%%%%%%%%%%%%%%%%%%%%%%%
\begin{equation}
\label{Eq.WC-Ionization}
\Delta E^{ion}_{n}=n\gamma \sqrt{N_{B}}\left[ \frac{3}{2}-\frac{3n}{8N_{B}}\right] +(k_{B}T_{0})l_{B}Z_{c}^{2}\frac{(n+1)n}{2a}\: .
\end{equation}
%%%%%%%%%%%%%%%%%%%%%%%%%%%%
% 
In Fig. 2 we compare the predictions of eqs. (\ref{Eq.oc} -- \ref{Eq.WC-Ionization})
to our simulation data, which shows excellent agreement. Our numerical data
for $ \Delta E^{ion}_{1} $ for $ N_{B}=15 $, 45, and 75, the value of
$ \Delta E^{OC}_{1} $ for $ N_{A}=90 $, as well as the corresponding values
for $ \gamma  $, which have been used for Fig.~2 can be found in Table~1.
They show that $ \gamma  $ is almost independent of $ N $. The value of
$ \gamma  $ can also be compared to the prediction of WC theory applied to
an infinite plane which leads to the value $ 1.96l_{B}Z_{c}^{2}\sqrt{\frac{1}{F}}\approx 2.76 $
\cite{Bonsall_PRB_1977}, where $ F $ denotes the surface area of the colloid.
The difference of 10 \% to WC theory is presumably related to the fact that
we do not deal with purely planar correlations but have a finite spherical geometry.

%%%%%%%%%%%%%%%%
%TABLE 1
\begin{table}
\caption{
Measured value, for an \textit{isolated} colloid, of the first ionization energy
\protect\protect$ \Delta E^{ion}_{1}\protect \protect $ for \protect\protect$ N_{B}=15\protect \protect $,
45, 75, and the energy gain for the first overcharging counterion \protect\protect$ \Delta E^{OC}_{1}\protect \protect $
for \protect\protect$ N_{A}=90\protect \protect $. The value of \protect$ \gamma \protect $
can be compared to the prediction of WC theory for an infinite plane, which
gives 2.76, compare text.
}
\label{tab:1}
\begin{center}
\begin{largetabular}{cccc}
\hline 
$Q/e$&
$N$&
$\Delta E_{1}/k_{B}T_{0}$&
$ \gamma /k_{B}T_{0} $\\
\hline 
-30&
15&
17.9 &
2.26 \\
 -90&
45&
29.2 &
2.42 \\
-150&
75&
37.4 &
2.50 \\
-180&
90&
-35.3&
2.47 \\
\hline 
\end{largetabular}
%{\small \par{}} \vspace*{-0.1cm}
\end{center}
\end{table}
%%%%%%%%%%%%%%%%%%%%%%%%%%%%%%%%

With the help of Eqs. (\ref{Eq.oc}, \ref{Eq.WC-Ionization}), one can try to
predict the curves of Fig. 1 for finite center-center separation \textit{R}.
Using for colloid $ A $ and $ B $ the measured values $ \gamma _{A} $
and $ \gamma _{B} $, we obtain for the electrostatic energy difference at
finite center-center separation $ R $
%
%%%%%%%%%%%%%%%%%%%%%%%
\begin{equation}
\label{eq.finiteR}
\Delta E_{n}(R)=\Delta E^{ion}_{n}+\Delta E^{OC}_{n}=
\frac{3}{2}n\gamma _{B}\sqrt{N_{B}}[1-\frac{n}{4N_{B}}]-
\frac{3}{2}n\gamma _{A}\sqrt{N_{A}}[1+\frac{n}{4N_{A}}]+
k_{B}T_{0}l_{B}Z_{C}^{2}\frac{n^{2}}{a}(1-\frac{a}{R})\: .
\end{equation}
%%%%%%%%%%%%%%%%%%%%%%%
%
The quality of the theoretical curves can be inspected in Fig. 1. The prediction
is is very good for large separations, but the discrepancies become larger for
smaller separations, and one observes that the actual simulated energies are
lower. With the help of Eq. (\ref{eq.finiteR}) we can establish a simple criterion,
valid for large macroionic separations, for the necessary charge asymmetry $ \alpha  $
to produce an ionized ground state of two unlike charged colloids with the same
size:
%
%%%%%%%%%%%%%%%%%%%%%%%
\begin{equation}
\label{Eq. criterion}
\frac{3}{2}\gamma \left( \sqrt{N_{A}}-\sqrt{N_{B}}\right) >\frac{(k_{B}T_{0})l_{B}Z_{c}^{2}}{a}\: .
\end{equation}
%%%%%%%%%%%%%%%%%%%%%%%
%
%%%%%%%%%%%%%%%%%%%%%
%FIG 2 AND FIG 3 GLUED
\begin{figure}[b]
\twofigures[width = 6.8 cm]{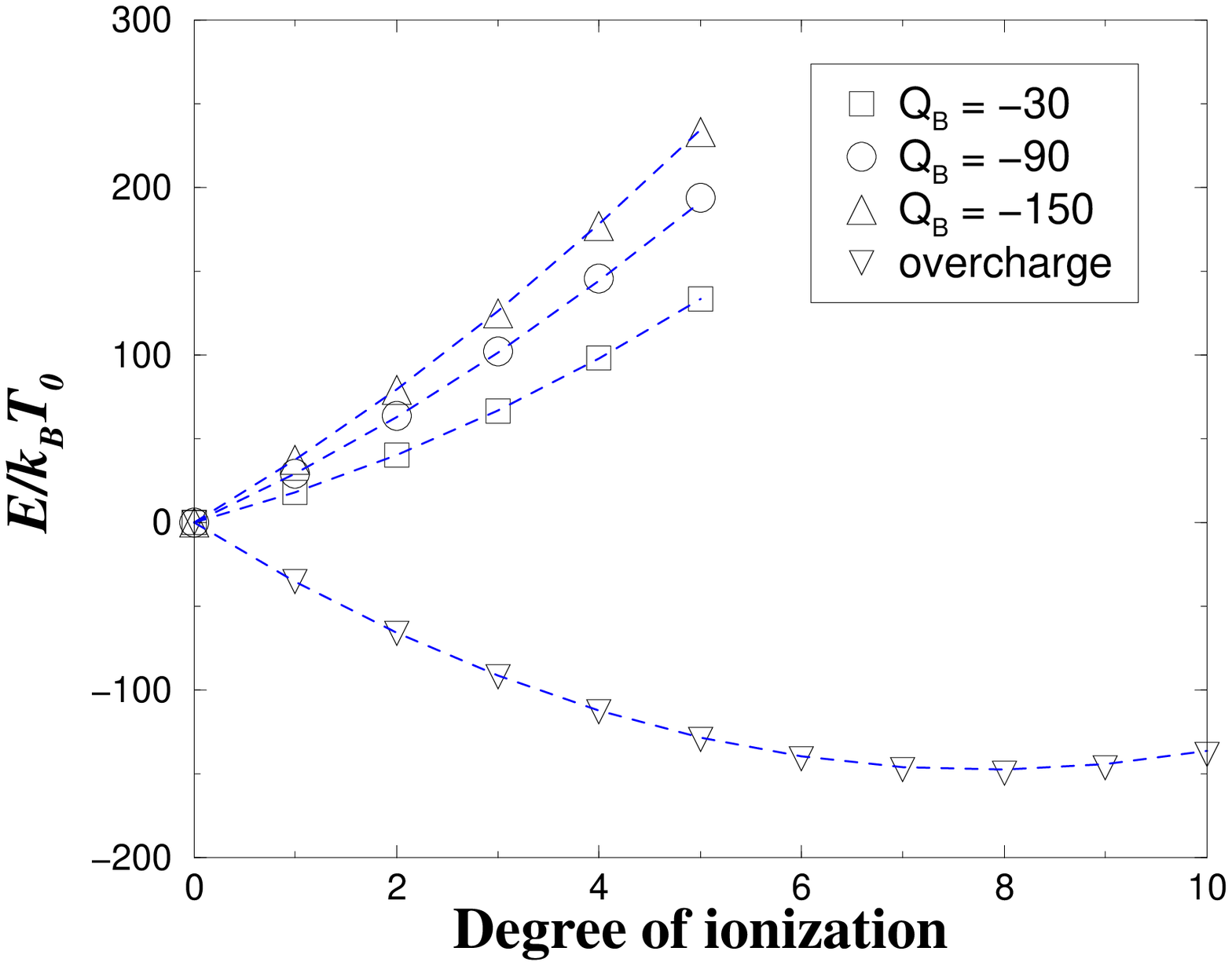}{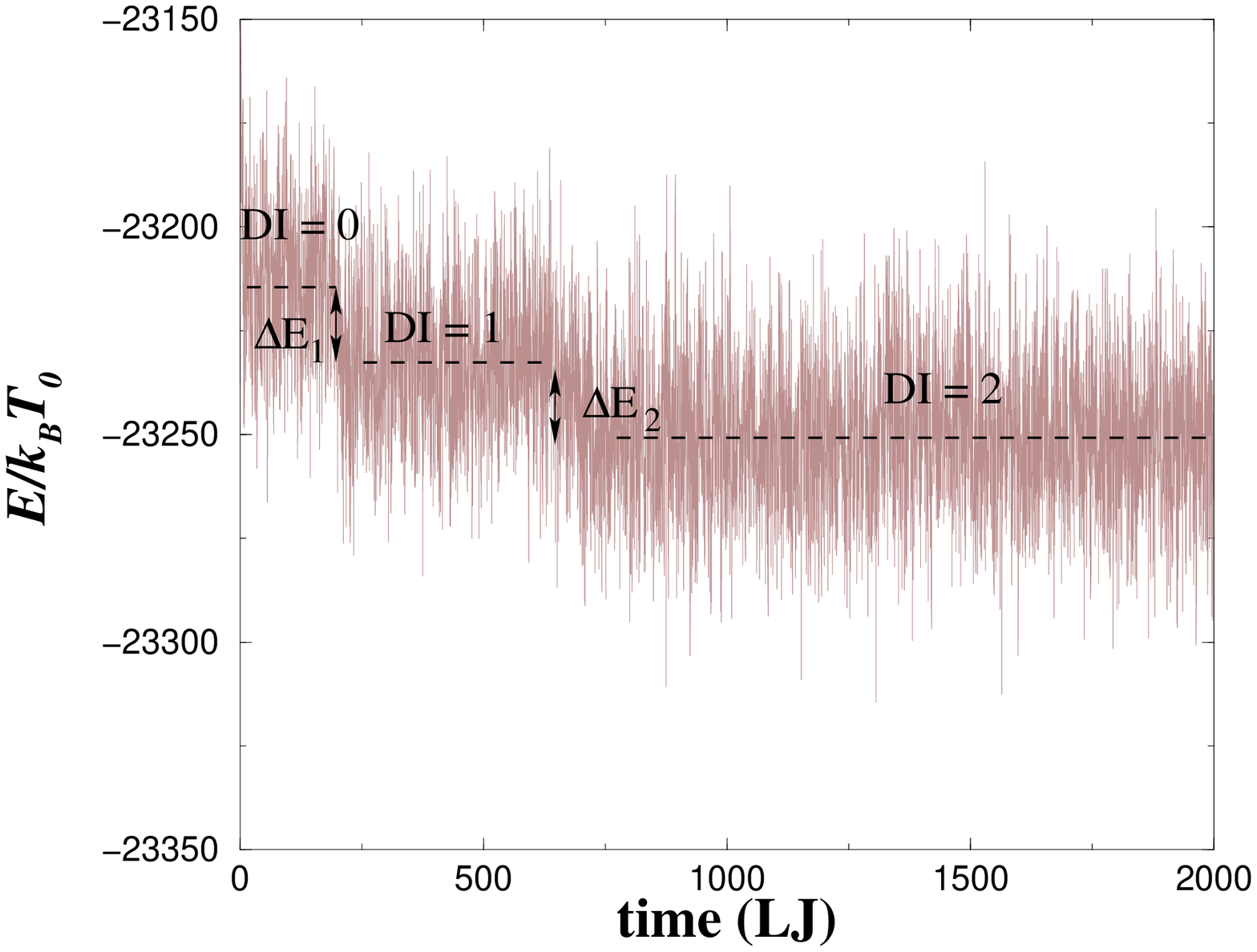}
%%fig 2
\caption{
Total electrostatic energy as a function of the degree of ionization
for zero temperature configurations of an \textit{isolated} colloid. The three
upper curves correspond to the ionization energy for the three typical charges
$ Q_{B}/e $ ($ -30,-90 $ and $ -150 $). The lower curve corresponds
to the energy gained by overcharging ($ Q_{A}/e=-180) $. Dashed lines were
obtained using eqs. (\ref{Eq.oc}, \ref{Eq.WC-Ionization}) with the measured
values for $ \gamma  $ from Table 1.
}
\label{ionization}
%%fig 3
\caption{
Relaxation, at room temperature $ T_{0}=298K $, of an initial neutral
state towards ionized state. Plotted is the total electrostatic energy versus
time (LJ units), for $ Z_{B}=30 $ and $ R/a=2.4 $. Dashed lines lines
represent the mean energy for each DI state. Each jump in energy corresponds
to a counterion transfer from the macroion \textit{B} to macroion \textit{A}
leading to an ionized state ($ DI=2 $) which is lower in energy than the
neutral one. The two energy jumps $ \Delta E_{1}/k_{B}T_{0}=-20 $ and $ \Delta E_{2}/k_{B}T_{0}=-17 $
are in very good agreement with those of Fig. 1c (-20.1 and -16.3).
}
\label{temperatute}
\end{figure}
%%%%%%%%%%%%%%%%%%%%%

The physical interpretation of this criterion is straightforward. The left term
represents the difference in correlation energy and the right term the monopole
penalty due to the ionization process. This means that the correlational energy
gained by overcharging the highly charged colloid \textit{A} must overcome the
loss of correlation energy as well as the monopole contribution (\textit{two}
penalties) involved in the ionization of colloid \textit{B}. If one uses the
parameters of the present study one finds the requirement $ N_{B}<66 $ to
get a stable ionized state. This is consistent with our findings where we show
in Fig.~1 that for $ N_{B}=75 $\textit{,} and \textit{R/a} = 4.25, no ionized
ground state exists whereas for $ N_{B}=60 $ we observed one even for infinite
separation. The criterion Eq. (\ref{Eq. criterion}) is merely a sufficient
condition, since we showed in Fig.~1 that when the colloids are close enough
this ionized state can appear even for smaller macroion charge asymmetry due
to enhanced intercolloidal correlations. If the colloids have the different
radii this can be can accounted for by simply replacing $ N_{i}^{1/2} $ by
the concentration of counterions $ (N_{i}/F_{i})^{1/2} $, and redefining
$ \gamma  $ Eq. (\ref{Eq. criterion}), in Eq. (\ref{Eq. criterion}).

At this stage, on looking at the results presented above, it appears natural
and straightforward to establish an analogy with the concept of ionic bonding.
It is well known in chemistry that the electro-negativity concept provides a
simple yet powerful way to predict the nature of the chemical bonding 
\cite{Pauling_1939}.
If one refers to the original definition of the electro-negativity given by
Pauling \cite{Pauling_1939}: ``the power of an
atom in a molecule to attract electrons to itself'', the role of the bare charge
asymmetry becomes obvious. Indeed, it has an equivalent role at the mesoscopic
scale as the electron affinity at the microscopic scale. Another interesting
analogy is the influence of the colloidal separation on the stability of the
ionized state. Like in diatomic molecules, the ionized state will be (very)
stable only for sufficiently short colloid separations. Nevertheless, one should
not push too far this analogy. Indeed, in many respects it breaks down, and
these are in fact important and interesting points. One concerns the existence
of an ionized ground state in colloidal system for \textit{large} colloid separation,
providing that $ \alpha  $ is large enough. In an atomistic system this is
impossible since even for the most favorable thermodynamical case, namely CsCl,
there is a cost in energy to transfer an electron from a cesium atom to a chlorine
atom. Indeed, the smallest existing ionization energy (for Cs, 376 kJ mol$ ^{-1} $)
is greater in magnitude than the largest existing electron affinity (for Cs,
349 kJ mol$ ^{-1} $). In other terms, for atoms separated by large distances
in the gas phase, electron transfer to form ions is always energetically unfavorable.

%%%%%%%%%%%%%%%%%%%
%FIG 4
\begin{figure}[t]
\onefigure[width = 8.0 cm]{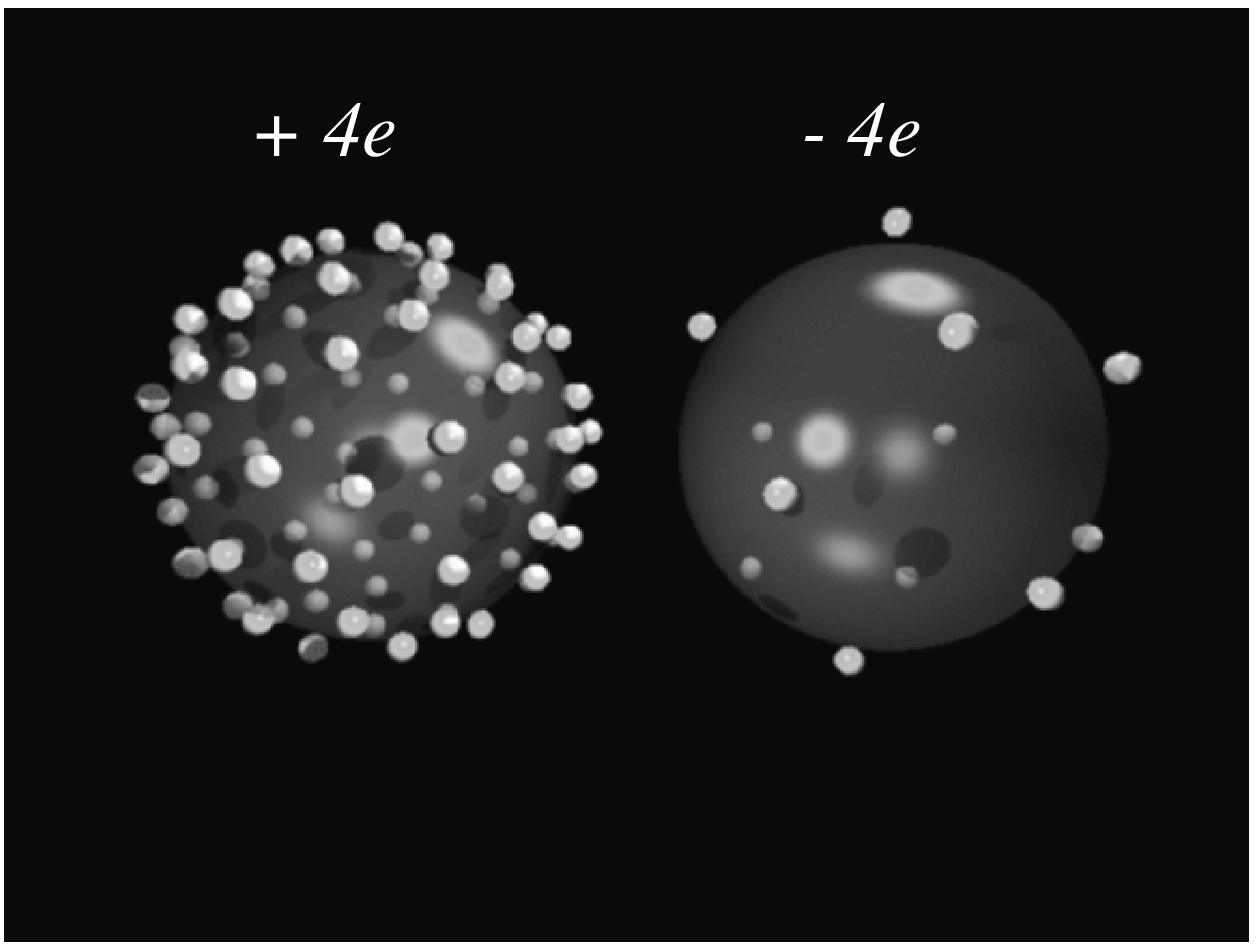}
\caption{
Snapshot of the ionized state ($ DI=2 $) obtained in the relaxation
process depicted in Fig. 3, with the net charges +4\textit{e} and -4\textit{e}
as indicated.
}
\label{snap}
\end{figure}
%%%%%%%%%%%%%%%%%%%%%
As a last result, aimed at experimental verification, we show that an ionized
state can also exist at \textit{room temperature} $ T_{0} $. Figure 3 shows
the time evolution of the electrostatic energy of a system $ Z_{A}=180 $
with $ Z_{B}=30 $, $ R/a=2.4 $ and $ f_{m}=7\cdot 10^{-3} $, where
the starting configuration is the neutral state (\textit{DI} = 0). One clearly
observes two jumps in energy, $ \Delta E_{1}=-19.5\, k_{B}T_{0} $ and $ \Delta E_{2}=-17.4\, k_{B}T_{0} $,
which corresponds each to a counterion transfer from colloid B to colloid A.
These values are consistent with the ones obtained for the ground state, which
are$ -20.1\, k_{B}T_{0} $ and $ -16.3\, k_{B}T_{0} $ respectively. Note
that this ionized state (\textit{DI} = 2) is more stable than the neutral but
is expected to be metastable, since it was shown previously that the most stable
ground state corresponds to \textit{DI} = 5. The other stable ionized states
for higher \textit{DI} are not accessible with reasonable computer time because
of the high energy barrier made up of the correlational term and the monopole
term which increases with \textit{DI} \cite{Messina_PRL_2000}. In Fig.
4 we display a typical snapshot of the ionized state (\textit{DI} = 2) of this
system at room temperature.

Obviously, these results are not expected by a DLVO theory even in the asymmetric
case (see e. g. \cite{DAguanno}). Previous simulations of
asymmetric (charge and size) spherical macroions \cite{Elshad_PRE_1998}
were also far away to predict such a phenomenon since the Coulomb coupling was
weak (water, monovalent counterions).

In summary, we have shown that the ground state of two unlike charged spherical
macroions is mainly governed by two important parameters, namely the bare charge
asymmetry $ \alpha  $ and the colloids separation \textit{R}. If $ \alpha  $
is high enough, the ground state corresponds to the so-called ionized state,
whatever the macroions separation \textit{R} is. In return, the degree of ionization
depends on \textit{R}. Furthermore, for large \textit{R}, we have established
a criterion for $ \alpha  $, allowing to predict when a stable ionized configuration
can be expected. The bare charge difference $ \alpha  $ plays an analogous
role to the electron affinity difference between two atoms forming a molecule
with ionic bonding. We demonstrated that the results presented here for the
ground state can lead to a stable ionic state even at room temperature providing
that the Coulomb coupling and/or the charge asymmetry is sufficiently large.
This is a possible mechanism which could lead to long range attractions, even
in bulk. Future work will treat the case where salt ions are present. Finally,
it would be desirable to theoretically quantify the influence of intercolloidal
correlations at short separations in a similar fashion as we have done for large
separations.

\stars 
This work is supported by \textit{Laboratoires Europ\'eens Associ\'es} (LEA). One
of the authors R. M. thanks E. Allahyarov for fruitful discussions. \newpage

% end of the MAIN TEXT
%\newpage
%

%
%\newpage
%

%\listoffigures{}
%\section{Figures and Tables}

%\begin{figure}


\begin{thebibliography}{10}
\bibitem{NOTE_MD}
Since we are using (MD) simulations,
to avoid barrier energy problem the counterions were generated at the vicinity
of the colloid surface.
\end{thebibliography}
\end{document}